\documentclass[aps,prb,showpacs,amsmath,amssymb,groupedaddress,preprint]{revtex4-1}

\usepackage{graphicx,hyperref}
\usepackage{multirow}

\newcommand{\tc}{$T_{\rm{C}}$}
\newcommand{\tb}{$T_{\rm{B}}$}

\begin{document}

\preprint{\today}
\title{Structural and Magnetic Characteristics of MnAs Nanoclusters Embedded in Be-doped GaAs}

\author{D. W. Rench}
\author{P. Schiffer}
\author{N. Samarth}
\affiliation{Department of Physics and Materials Research Institute, The Pennsylvania State University, University Park, Pennsylvania  16802, USA}

\begin{abstract}
We describe a systematic study of the synthesis, microstructure  and magnetization of hybrid ferromagnet-semiconductor nanomaterials comprised of MnAs nanoclusters embedded in a p-doped GaAs matrix. These samples are created during the \textit{in situ} annealing of Be-doped (Ga,Mn)As heterostructures grown by molecular beam epitaxy. Transmission electron microscopy and magnetometry studies reveal two distinct classes of nanoclustered samples whose structural and magnetic properties depend on the Mn content of the initial (Ga,Mn)As layer. For Mn content in the range $5\% - 7.5 \%$, annealing creates a superparamagnetic material with a uniform distribution of small clusters (diameter $\sim 6$ nm) and with a low blocking temperature (\tb $\sim 10$ K). While transmission electron microscopy cannot definitively identify the composition and crystalline phase of these small clusters, our experimental data suggest that they may be comprised of either zinc-blende MnAs or Mn-rich regions of (Ga,Mn)As. At higher Mn content ($\gtrsim 8 \%$), we find that annealing results in an inhomogeneous distribution of both small clusters as well as much larger NiAs-phase MnAs clusters (diameter $\sim 25$ nm). These samples also exhibit supermagnetism, albeit with substantially larger magnetic moments and coercive fields, and blocking temperatures well above room temperature.

\end{abstract}
\maketitle

\section{Introduction}
\label{sec:intro}

Integrating ferromagnetism with semiconductors is a central theme in semiconductor spintronics as it provides a potential route towards spin-based semiconductor devices. \cite{Awschalom:2007vq} Significant effort has been invested in this context in understanding and improving the properties of the ferromagnetic semiconductor (Ga,Mn)As. \cite{Macdonald:2005ew,Dietl:2010ve}  Despite remarkable advances in recent years that include an enhancement of the Curie temperature to about 195 K, \cite{Khazen:2010cr} as well as the demonstration of novel functionality such as electrically modulated magnetization,\cite{Chiba:2008rt} the technological potential of (Ga,Mn)As still remains elusive. However, the material 
provides an excellent model system for testing new semiconductor spintronic device concepts.\cite{Zhu:2007kt,Zhu:2008ck,Endo:2010zr}  Many researchers have undertaken a parallel search for alternative ferromagnetic semiconductors with a high \tc,\cite{Dietl:2010ve} but this has been stymied by a lack of materials with significant remanent magnetization under ambient conditions. Yet another approach to integrating ferromagnetism with semiconductors involves ``hybrid'' materials that combine phase-separated metallic ferromagnets with a semiconductor ``host'' \cite{Dietl:2010ve,Kuroda:2007dq,Bonanni:2008bh}. Such materials could hypothetically take advantage of a higher \tc~within regions of above-room-temperature ferromagnetic metals dispersed throughout an electrically-controllable semiconductor. 

One such candidate material consists of MnAs nanoclusters embedded inside a GaAs host matrix.\cite{DeBoeck:1996zr} Though long regarded as an impediment to achieving carrier-controlled ferromagnetism in homogeneous (Ga,Mn)As, interest in this clustered MnAs:GaAs system has been resurrected because of recent demonstrations of novel semiconductor spintronic devices that utilize spin-dependent interactions between carriers in extended states of the host semiconductor and vicinal MnAs nanostructures.\cite{Hai:2009dz,Hai:2010uq}

In this paper, we focus on developing a hybrid material wherein MnAs clusters are embedded in a conducting p-doped GaAs host lattice. The eventual goal is to realize a hybrid system wherein itinerant charge carriers in a doped semiconductor interact with a distribution of nanoscale ferromagnets producing useful magnetoelectronic effects at room temperature. We note that earlier studies of such p-GaAs:MnAs hybrid materials have been reported, \cite{Yu:2006bs} but our present studies reveal distinctly different crystal structures and magnetic properties.  In addition, we report a systematic set of  measurements that examine the various co-dependent physical properties of this hybrid system.
 
Earlier work has identified two principal approaches for designing hybrid MnAs:GaAs clustered systems: the annealing of (Ga,Mn)As thin films either post-growth\cite{Kwiatkowski:2007hc} or \textit{in situ}. \cite{Hai:2007ij} Both approaches have their advantages and disadvantages.  Post-growth annealing is quick and allows for many different annealing recipes to be attempted in a short time period.  \textit{In situ} annealing is much more controlled and allows for the annealing to occur in a high vacuum, As-rich environment.  Our work focuses on the latter approach.  For this study, we varied the Be and Mn doping levels methodically and explored (to a lesser extent) variations in annealing times.  Detailed magnetization measurements were performed to determine the nature of the temperature- and magnetic-field-dependence of these samples.  Two types of cluster materials were discerned based on vastly different magnetic properties and superparamagnetic blocking temperatures as determined by different magnetometry techniques. Also, the size distribution and crystal orientations of the different cluster types were evaluated using high resolution cross-sectional tunneling electron microscopy (HRXTEM).  Finally, the size distribution was verified to be log-normal, as has been observed in other studies of this material type. \cite{Kwiatkowski:2009ys}

\section{Experimental Details}
\subsection{MBE Growth}
\label{sec:MBE}

\begin{figure}
\includegraphics[scale=0.65]{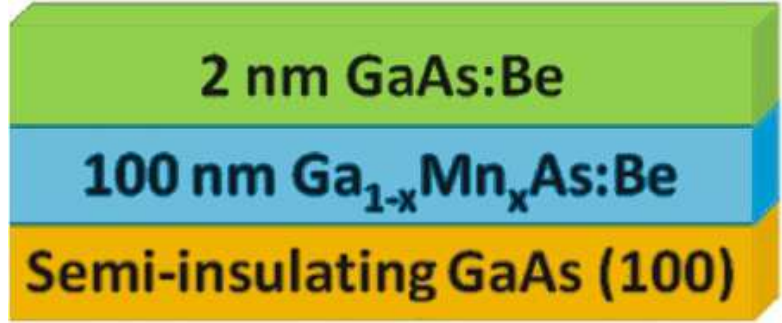}
\caption{(Color online) Schematic showing the structure of our sample growths.  The (Ga,Mn)As layer becomes a hybrid GaAs:MnAs cluster layer upon the growth of the GaAs:Be capping layer, which is effectively a high-temperature thermal annealing step.}
\label{fig:SampleStructure}
\end{figure}

All the samples in this study were grown via molecular beam epitaxy (MBE) on semi-insulating (100) GaAs substrates.  All substrates had their oxides desorbed at 600 $^{\circ}$C, followed by the growth of an approximately 170 nm thick semi-insulating buffer layer to smooth out the interface between the substrate and sample layers.  Then, a 100 nm layer of $\text{Ga}_{1-x}\text{Mn}_x\text{As:Be}$ was grown on the buffer layer at approximately 245 $^{\circ}$C, where the Mn content was varied for different samples.  To complete the material, a 2 nm high-temperature GaAs:Be layer was grown as a cap, with the Be content of the cap corresponding to a 50 $^{\circ}$C decrease in the Be cell temperature.  This was done in an effort to roughly equalize the hole concentration between the magnetically active layer and the capped layer.  Since As antisites in the low-temperature (Ga,Mn)As layer would have had a double-donor compensating effect on the holes provided by the Be dopant, the material required a larger amount of Be to compensate (the Mn would have an insignificant effect on carrier concentration once it was forced into MnAs nanoclusters by the annealing process).  The final material produced was a p-doped GaAs matrix on a semi-insulating substrate which had MnAs nanoscale clusters embedded within it.  We note that the capping layer served a three-fold purpose.  First, we found through our own extensive post-growth annealing that a GaAs cap allowed for a system in which Mn could not anneal out of the sample.  Second, the presence of such a cap provided additional insight into the cluster formation process: as discussed later in this manuscript, HRXTEM studies show that the clusters form within the capping layer, but not in the buffer layer, which is an important consideration in developing these hybrid materials.  Third, the growth of this high-temperature cap acted as the thermal annealing process which formed the (Ga,Mn)As layer into a cluster material.  In order to grow this thin cap, we increased the substrate temperature to a standard high-temperature GaAs growth temperature of 600 $^{\circ}$C at 20 $^{\circ}$C/min and grew the film for approximately 22 seconds.  Other growths were carried out for more than 15 minutes to test the effects of a longer anneal time; the resulting films exhibited similar magnetic behavior as those with shorter annealing times.  Therefore, the relatively slow ramp rate was all that was necessary to form nanoclusters, and a longer annealing time at the final temperature had no effect on the samples' magnetic behavior.  A diagram of the sample structure is shown in Figure \ref{fig:SampleStructure}.

We performed a systematic study of the magnetic properties of this material, with principal focus on effects of varying the Mn content of the samples.  While we also studied samples in which the Be content was varied, these did not result in any significant changes to the magnetization.  A summary of the samples grown at the time of publication is given in Table \ref{table:SampleTable}.  

We grew five principal samples where the Mn content was changed by approximately 0.65\% for each sample. The Mn content quoted for all the samples quoted within this work is an estimate based upon measurements of the molecular beam flux determined by the beam equivalent pressure ratios prior to the sample growth, and as such is not an accurately calibrated value. It is nonetheless a reasonable estimate since we have over the years developed a correlation between the growth parameters used in our MBE system and post-growth analysis of the actual Mn content using secondary ion mass spectroscopy. To determine carrier concentrations in our samples, we carried out Hall effect measurements using a standard Hall bar geometry and found hole concentrations of $p \sim 10^{18}$ cm$^{-3}$.  This value is reasonable based on previous calibration measurements of both (Ga,Mn)As and Be-doped GaAs in our MBE system, as well as expected compensation effects.  This hole concentration likely corresponds to the doping level of the GaAs matrix that envelopes the MnAs clusters; however, we caution that in such a composite material, a proper analysis of electrical transport in general (and the Hall effect in particular) is likely to be complicated.

\subsection{SQUID Magnetometry}
\label{sec:SQUID}

\begin{figure}
\includegraphics[scale=0.65]{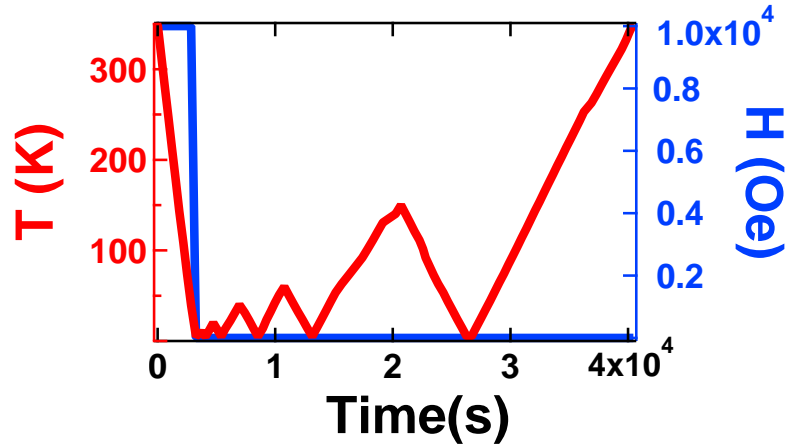}
\caption{(Color online) Description of thermoremanent magnetization measurements.  Samples are field-cooled in a 1 Tesla field from 350 K to 5 K, then a field of 0.5 Oe is applied while measuring magnetization from 5 K to different final temperatures, returning to 5 K every time a new final temperature has been reached.}
\label{fig:TRExplain}
\end{figure}

A variety of magnetization techniques were employed for the characterization of the magnetic properties of our samples.  We performed magnetic field sweeps, temperature sweeps (both field-cooled and zero-field-cooled), and thermoremanent magnetization measurements.   The field sweeps are performed by putting the sample at a set temperature and then applying a saturating magnetic field $+H$.  Measurements of the magnetization are then taken from $+H$ to $-H$ and back to $+H$ to observe any hysteretic effects.  The field-cooled (FC) measurements were obtained after cooling the sample from a temperature $T = 350$ K (the highest temperature our SQUID can safely achieve) to $T = 5$ K in a saturating field.  The sample chamber and magnet are then carefully demagnetized before beginning these measurements due to the relatively small signal generated by these materials.  Once the sample temperature was stabilized at 5 K, we applied a 50 Oe measuring field and took data from $T = 5$ K up to $T = 350$ K. The zero-field-cooled measurements (ZFC) are very similar to the FC measurements except that the sample is cooled in a zero field environment instead of a saturating field. The thermoremanent magnetization measurements are performed by cooling the sample to $T = 5$ K in a saturating magnetic field. The measuring field applied is carefully brought to a value of $0 < H < 1$ Oe instead of the standard 50 Oe. It should be noted that such a small field can only be accurately attained if a careful demagnetization of the sample chamber and the magnet is performed to remove any remanent field. The magnetization is then measured while the sample is heated by an amount $\Delta T_1$ and subsequently cooled back to $T = 5$ K. We then continue measuring the magnetization of the sample as it is heated by an amount $\Delta T_2 > \Delta T_1$ and again cooling to $T = 5$ K. This process is repeated over and over again with successively larger $\Delta T_i$ until the sample has been brought all the way to $T = 350$ K. This procedure is illustrated in Figure \ref{fig:TRExplain}. Also, in all the data presented here, the diamagnetic contribution from the substrate is always subtracted by measuring the magnetization signal at high field where the diamagnetic signal from the substrate dominates over the saturated signal from the ferromagnetic material. The diamagnetic background signal is then modeled as a linear fit and subtracted.

\begin{figure}
\includegraphics[scale=0.65]{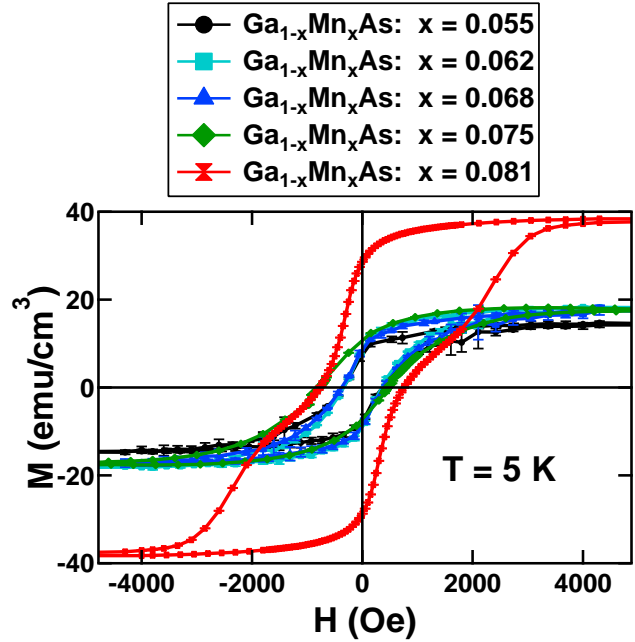}
\caption{(Color online) Comparison of the M(H) data for samples with differing Mn concentrations.  The clear difference between the measurements for the four lower-concentration samples and the most highly concentrated sample indicates an abrupt change in the character of the material.  Note that the Mn concentrations are calculated from the Ga:Mn flux ratios immediately prior to sample growth.}
\label{fig:MnVariedHyst}
\end{figure}

Figure \ref{fig:MnVariedHyst} shows the magnetization as a function of applied magnetic field for the five samples of interest, revealing a substantial difference in the magnetic characteristics of samples with Mn content in the range of 5\% to 7.5\% and that of a sample with a Mn content of 8.1\%.  The behavior changes qualitatively at around 8\% Mn concentration, suggesting two types of materials.  For the purposes of this manuscript, we classify these two different material systems as ``Type I'' (for the low-Mn growths) and ``Type II'' (for the high-Mn growths).  More samples were grown at 8\% and higher content, verifying the reproducibility of this Type II material.

\begin{figure}
\includegraphics[scale=0.5]{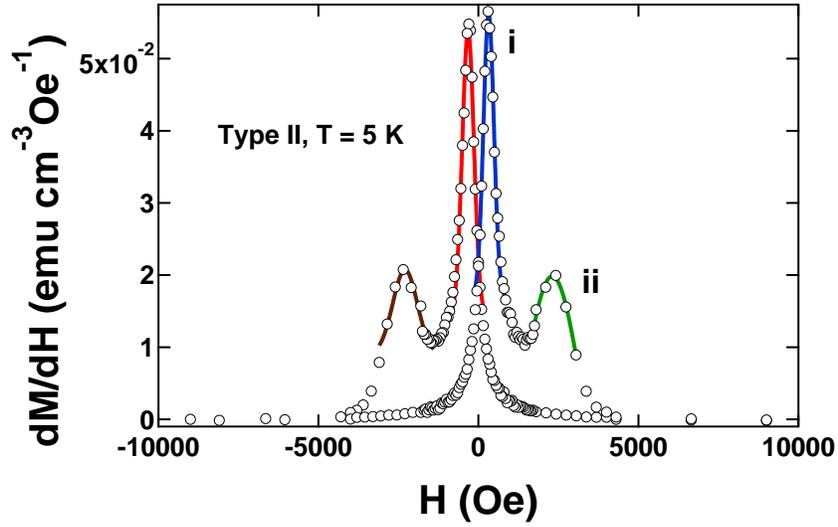}
\caption{(Color online) Gaussian fits of the derivative with respect to magnetic field of the magnetization for the 8.1\% Mn sample in Figure \ref{fig:MnVariedHyst}.  The fitting determined that, in this Type II sample, the small clusters had a coercivity of (i)322 Oe and the large ones a coercivity of (ii)2300 Oe.}
\label{fig:dMdH}
\end{figure}

There are two other important observations to make about these data:  the 7.5\% Mn sample shows exchange biasing and the 8.1\% Mn sample exhibits a "`bowing in"' of the hysteresis loop.  We attribute the exchange biasing to the sample being near the boundary in parameter space separating the Type I and Type II materials systems.  This sample has a low enough Mn content to behave like a Type I system with a significantly weaker saturation magnetization in comparison to a Type II system.  However, it has a high enough Mn content that the clusters discussed below are surrounded by a lightly-Mn-doped matrix of (Ga,Mn)As, which has the effect of biasing the clusters. This conclusion is supported by energy-dispersive x-ray spectroscopy (EDS) performed via scanning transmission electron microscopy (STEM) on the 7.5\% Mn sample.  We find a Mn content of approximately 6.6\% relative to the Ga in the matrix surrounding the small clusters in this sample. The area measured for this value contained no clusters; however, it should be noted that the Mn concentration appeared to be non-uniform throughout the sample, even where cluster formation did not occur.  With regard to the ``bowing in'' effect shown in Figure \ref{fig:MnVariedHyst}, we will show evidence in Section \ref{sec:structural} that this most likely arises from the coexistence of two unique MnAs-based cluster types contributing to the hysteresis.  By assuming two magnetically dissimilar cluster types, we can treat the hysteresis data as the sum of two hystereses:  one that exhibits a low saturation magnetization and a low coercivity and one that exhibits a high saturation magnetization and a high coercivity.  This hypothesis is further supported by plotting $dM/dH$ versus $H$ for the 8.1\% Mn sample and then fitting the resulting peaks to Gaussians, as shown in Figure \ref{fig:dMdH}; the peaks indicate unique coercivities of 322 Oe and 2300 Oe (i and ii in Figure \ref{fig:dMdH}, respectively).

Based on our magnetometry data, we have determined these samples to be superparamagnetic materials.  In considering the samples as superparamagnetic ensembles, we treat each ferromagnetic cluster to be single-domain, therefore acting as a ``superspin'' at temperatures well below the ferromagnetic transition temperature of the material within the cluster.  The energetic barrier to flipping the moment of the superspins is given by $E_{\rm{anis}} = K_{\rm{eff}}V$, where $K_{\rm{eff}}$ is the effective magnetic anisotropy constant and $V$ is the cluster volume; in these materials, the most significant contribution to $K_{\rm{eff}}$ is presumably the magnetocrystalline anisotropy, given the roughly spherical morphology indicated by the TEM measurements discussed in the next section. Below a characteristic blocking temperature, \tb, thermal fluctuations are insufficient to overcome $E_{\rm{anis}}$ and thus cannot flip the moment of the ferromagnetic clusters.  Therefore the moment of the samples for $T<$ \tb~ is determined by the magnetic history of the sample (i.e. in what magnetic field the sample was cooled from above \tb).  For $T>$ \tb, thermal fluctuations allow the superspin moments to fluctuate freely, and thus the magnetic behavior of the sample is analogous to that of a paramagnet.

\begin{figure}
\includegraphics[scale=0.5]{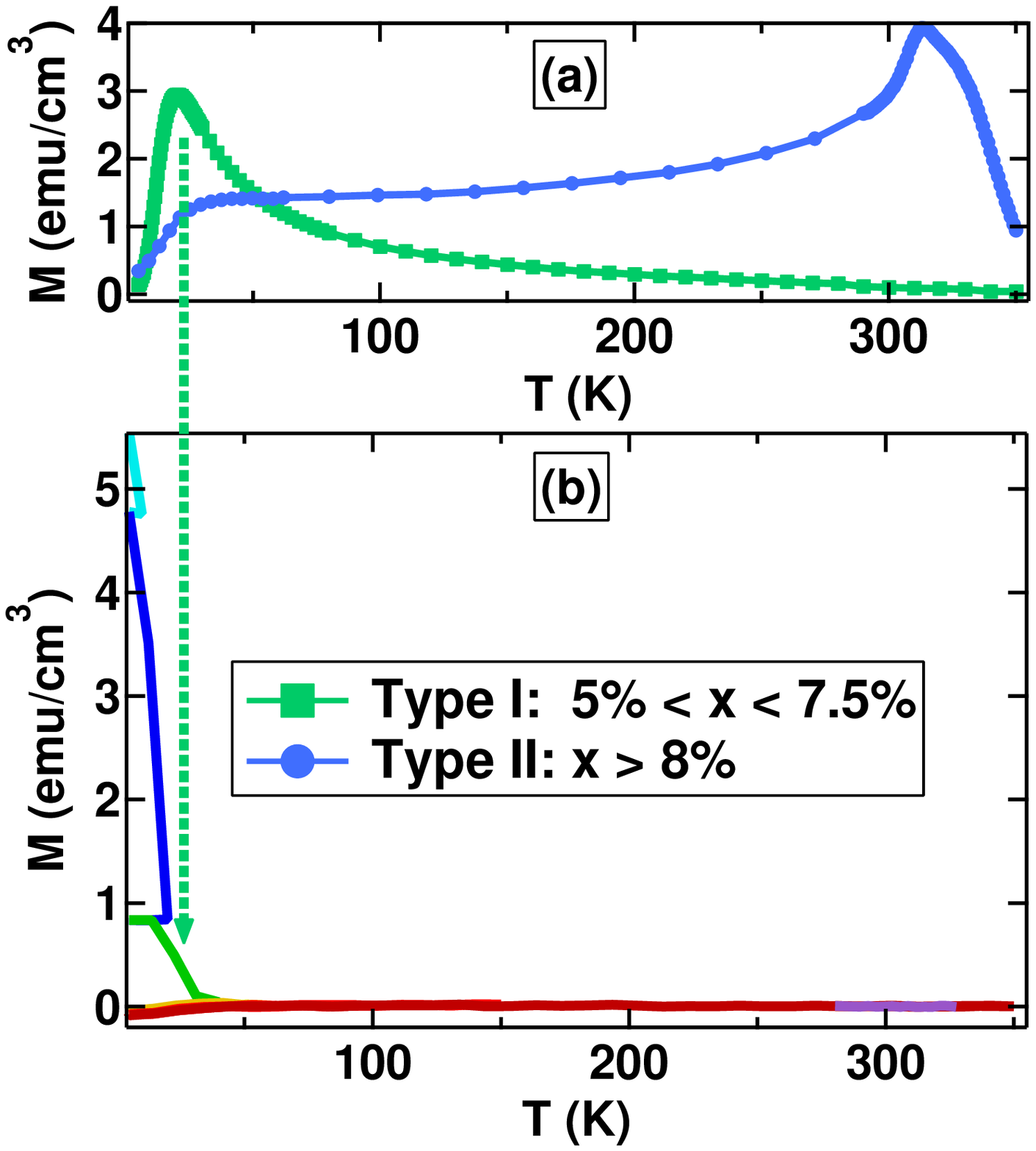}
\caption{(Color online) Zero-field-cooled (ZFC) measurement of magnetization for a Type I sample and a Type II sample as well as a thermoremanent magnetization measurement of the Type I cluster material.  The arrows indicate where transition points in the ZFC measurement correspond to features in the corresponding thermoremanent scan for the Type I sample.}
\label{fig:TempDataCompare_1}
\end{figure}

\begin{figure}
\includegraphics[scale=0.5]{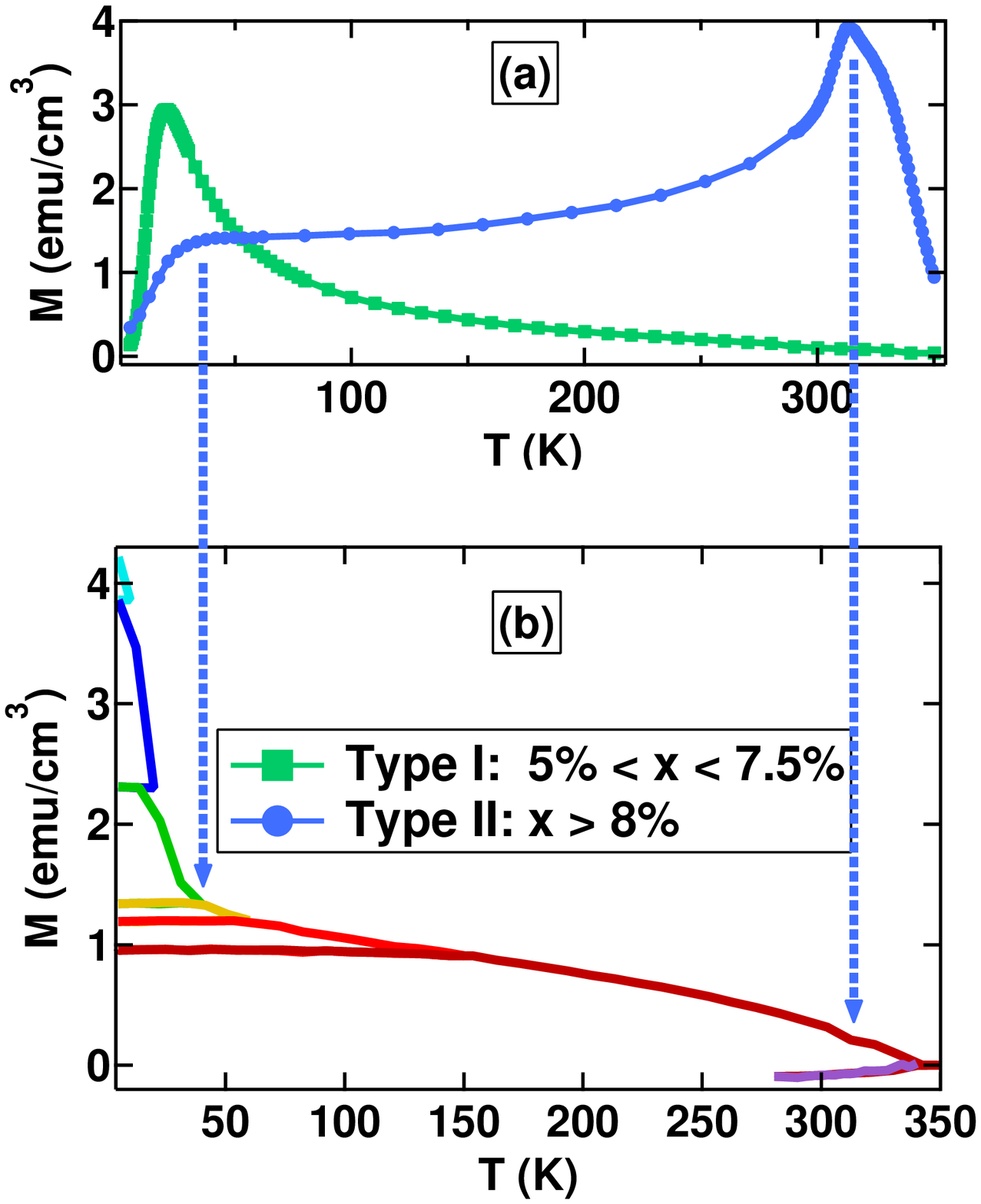}
\caption{(Color online) Zero-field-cooled (ZFC) measurement of magnetization for a Type I sample and a Type II sample as well as a thermoremanent magnetization measurement of the Type II cluster material.  The arrows indicate where transition points in the ZFC measurement correspond to features in the corresponding thermoremanent scan for the Type II sample.}
\label{fig:TempDataCompare_2}
\end{figure}

To better understand the differences in magnetization relaxation between Type I and Type II materials, we performed extensive measurements of the temperature dependence of the magnetization of our samples.  Figure \ref{fig:TempDataCompare_1} shows the zero-field-cooled (ZFC) measurements for a Type I and Type II sample as well as the thermoremanent magnetization\cite{Sawicki:2010qf} measurements for a Type I sample.  Figure \ref{fig:TempDataCompare_2} shows the same information for a Type II sample. If the superparamagnetic blocking temperature is exceeded in the presence of a very small measuring field, the clusters' moments do not align in any preferred direction, producing a net zero magnetization.  Thus, the temperature at which the magnetic measurements become repeatable on successive cycles is the effective blocking temperature.  Such measurements allow the characterization of superparamagnetic behavior in ensembles of ferromagnetic particles\cite{Sawicki:2010qf}, such as the clusters in our Type I and Type II samples.

\begin{figure}
\includegraphics[scale=0.5]{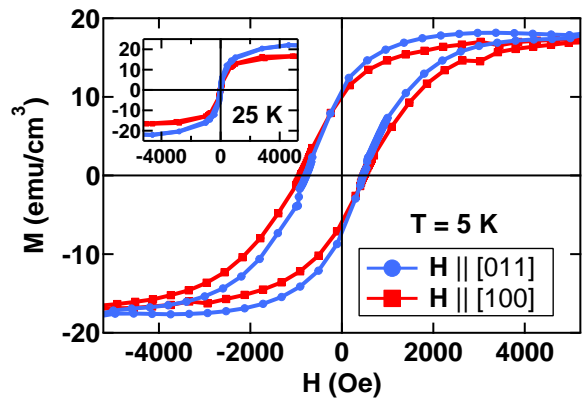}
\caption{(Color online) Main graph:  plot of M(H) at 5 K for two measurement orientations of a Type I sample.  Inset:  plot of M(H) at 25 K for two measurement orientations of a Type I sample.  At each temperature, the data for the two orthogonal orientations appear to vary only slightly.}
\label{fig:TypeIHyst}
\end{figure}

Using the data shown in Figure \ref{fig:TempDataCompare_1}, we found the blocking temperature of the 7.5\% Mn sample to be around 10 K; thus we took measurements of $M(H)$ for this sample at 5 K and 25 K, so that we could see the magnetic field dependence of the material both above and below its superparamagnetic transition point.  This data is presented in  Figure \ref{fig:TypeIHyst} which shows that the Type I material loses any resemblance to a standard ferromagnetic hysteresis curve, once the sample goes above 25 K.  This is reasonable considering the transitions we observed in our ZFC and thermoremanent data.  Also, Figure \ref{fig:TypeIHyst} shows a comparison at each temperature of two orthogonal measurement directions.  We observe only a slight difference in the saturation magnetization of the two orientations at 25 K.  The qualitative nature of the measurements does not seem to change in these two orientations.  

\begin{figure}
\includegraphics[scale=0.5]{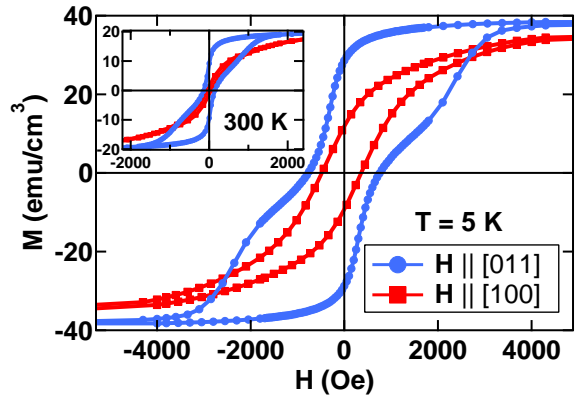}
\caption{(Color online) Main graph:  plot of M(H) at 5 K for two measurement orientations of a Type II sample.  Inset:  plot of M(H) at 300 K for two measurement orientations of a Type II sample.  The character of the plots at each temperature appear similar, and there is a substantial change due to the orientation of the sample.}
\label{fig:TypeIIHyst}
\end{figure}

Figure \ref{fig:TypeIIHyst} shows $M(H)$ measurements for the 8.1\% Mn Type II sample material and are performed at 5 K and 300 K. We find qualitatively different behavior and we display results at 300 K to demonstrate how the material would behave at the technologically-interesting regime of room temperature.  Interestingly enough, while the results seem to be relatively consistent for each orientation at 5 K compared to the same orientation at 300 K, the orientations themselves are vastly different.  This suggests that there is a strong crystalline dependence on the magnetic anisotropy for a Type II material, in sharp contrast to the data for the Type I material.  This may indicate that the Type I material forms magnetic clusters with random crystallographic orientations relative to the GaAs matrix but that the Type II material is constrained to orient its clusters in a specific direction relative to the GaAs matrix.  Our structural data discussed below supports that possibility in the Type II material, although the same could not be verified for the Type I material.

\subsection{Structural Study}
\label{sec:structural}
The magnetometry data presented in Section \ref{sec:SQUID} strongly suggest the formation of a cluster-based magnetic system due to the annealing of a (Ga,Mn)As film. This hypothesis is confirmed by high resolution TEM imaging.  Ideally, one would like to obtain both a real space image and a diffraction image for both the Type I and Type II materials.  Unfortunately, our beam resolution was not sufficient to focus on single clusters in the Type I system due to their extremely small size.  We obtained a nanobeam diffraction image that indicated a zinc blende structure, as one might expect for small MnAs:Ga clusters; \cite{Yokoyama:2005la} however, the lattice constants derived were also very close to those for GaAs, leading us to believe that we may have only obtained significant diffraction data from the surrounding GaAs matrix.  This is reasonable, considering the huge volume that the GaAs comprises in comparison to these small clusters.  

\begin{figure}
\includegraphics[scale=0.75]{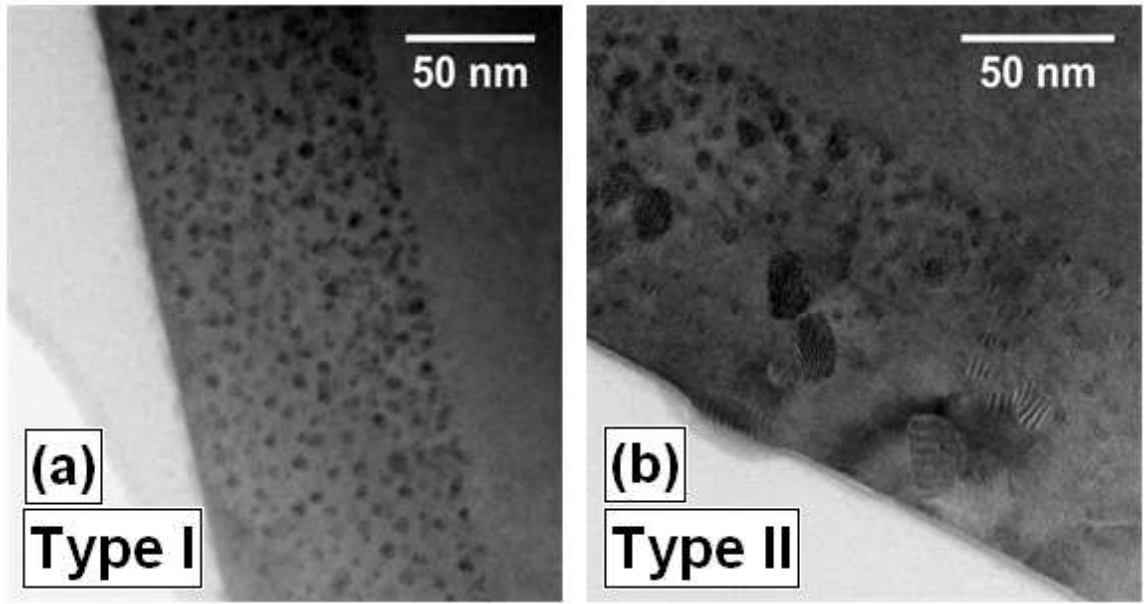}
\caption{(a) Real space HRXTEM image of the 7.5\% Mn Type I sample.  The sample appears to only be comprised of small nanoclusters, with a diameter on the order of 6 nm.  (b) Real space HRXTEM image of the 8.1\% Mn Type II sample.  This material appears to be a hybrid of small and large nanoclusters (the large ones having a diameter on the order of 15-25 nm).}
\label{fig:TEMCompare}
\end{figure}

\begin{figure}
\includegraphics[scale=0.75]{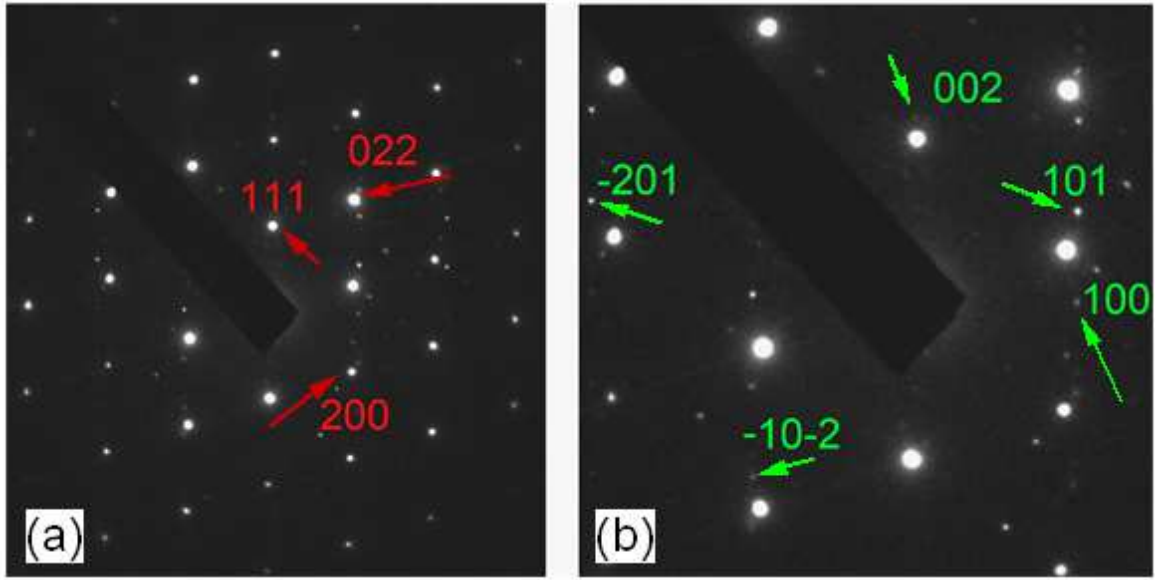}
\caption{(Color online) (a) TEM diffraction image of a large cluster in the Type II sample with the GaAs reflections indexed.  (b) Slightly magnified copy of (a) but with the MnAs reflections indexed.  This result corresponds to a NiAs-type structure with lattice constants of $a=3.73$ \AA~ and $c=5.76$ \AA, values which are very close to those for bulk MnAs as reported in the literature \cite{WILSON:1964kl}.}
\label{fig:TypeIIDiff}
\end{figure}

\begin{figure}
\includegraphics[scale=0.65]{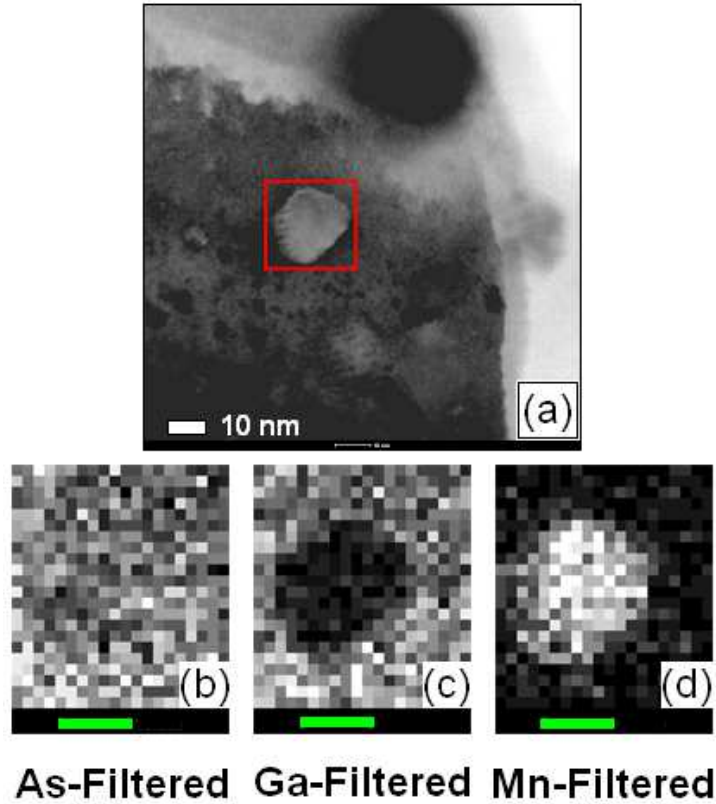}
\caption{Elemental mapping performed via EDS and scanning TEM (STEM) of the Type II sample, 203A.  (a) Real space image of the region mapped, with the square indicating the exact scanning area.  (b) Image of the scanned region filtered to only show the arsenic-rich regions.  (c) Image of the scanned region filtered to only show the gallium-rich regions.  (d) Image of the scanned region filtered to only show the manganese-rich regions.  The scale bars in (b), (c), and (d) are each equivalent to 10 nm.}
\label{fig:TypeIIEDS}
\end{figure}

While diffraction image results were unobtainable for the Type I material, we were able to obtain real space image data for both material systems, as is shown in Figure \ref{fig:TEMCompare}. This figure provides insight into the magnetometry data presented earlier in this paper.  In Figure \ref{fig:MnVariedHyst}, we noted that the shape of the hysteresis curve for the Type II sample (8.1\% Mn in the figure) was likely due to a coexistence of different cluster types.  These high-contrast images in Figure \ref{fig:TEMCompare} show the expected Moire fringes, and they also show obvious structural deviations in the large nanoclusters in comparison to the surrounding GaAs matrix.  We employed diffraction imaging and EDS in the Type II sample shown here to verify the crystal structure of these large clusters and determine how similar or dissimilar they are to the surrounding semiconductor material.  Figures \ref{fig:TypeIIDiff} and \ref{fig:TypeIIEDS} show the diffraction measurement and EDS elemental mapping (respectively) for a large nanocluster in a Type II sample. The image shows the same diffraction pattern twice.  Figure \ref{fig:TypeIIDiff}(a) has been indexed to show the various GaAs reflections present in these data.  Figure \ref{fig:TypeIIDiff}(b) is the same image but magnified slightly and has MnAs reflections indexed.  Since the cluster appears to exhibit both GaAs and MnAs reflections, it might lead one to believe that there is Ga present inside these larger clusters in the Type II samples.  However, Figure \ref{fig:TypeIIEDS} shows clearly that Ga is not present in any substantial quantity in comparison to Mn.  The reason for these GaAs reflections is simply that the sample measured was not perfectly thinned to the point at which the only material in the path of the electron beam was cluster material.  The beam traveled through GaAs on either side of the cluster during the measurement, causing the GaAs reflections to appear in the results.

\section{Discussion of Results}
\label{sec:discussion}

As we mentioned earlier, Figure \ref{fig:TypeIIEDS} clearly shows the elemental makeup of the large clusters.  Together with the diffraction measurements, this allows us to infer that the large clusters are comprised of pure MnAs in the NiAs phase embedded in the zinc-blende p-GaAs matrix.  The EDS results also show us that a clean and abrupt interface exists between the large clusters and the semiconductor material, which will be an important point of consideration for future electrical transport studies of this material.  This chemical information, combined with the structural and magnetic data already presented, support the hypothesis that the smaller clusters are of a different crystal structure from the larger ones. These smaller clusters may even have randomly oriented crystal directions or a zinc blende structure as has been suggested earlier. \cite{Yokoyama:2005la}

It is also interesting to note the regions in which the clustering occurs.  In Figure \ref{fig:TEMCompare}(a) we note that clustering occurs throughout the doped GaAs:Be cap layer and what was, upon growth, the (Ga,Mn)As:Be layer. However, the clusters are not present in the undoped buffer layer.  Also, in Figure \ref{fig:TEMCompare}(b) we see that the larger clusters occur near and upon the sample surface and that the smaller clusters form near the interface between the magnetic layer and the undoped buffer layer.  We hypothesize, based on these observations, that Be-doping facilitates Mn clustering.  A possible explanation for this is based on observations of the behavior of Mn interstitials when annealed at low temperatures in (Ga,Mn)As. \cite{Yu:2004fu} The Fermi level of an undoped region is known to be higher than that of a p-type region.  This difference in Fermi levels across the interface of the buffer and the magnetic layer may form an energy barrier to the Mn clusters that causes them to only form in Be-doped regions.  In order to verify this hypothesis, TEM studies will be performed on control samples that are entirely undoped and entirely doped in the near future.

\begin{figure}
\includegraphics[scale=0.5]{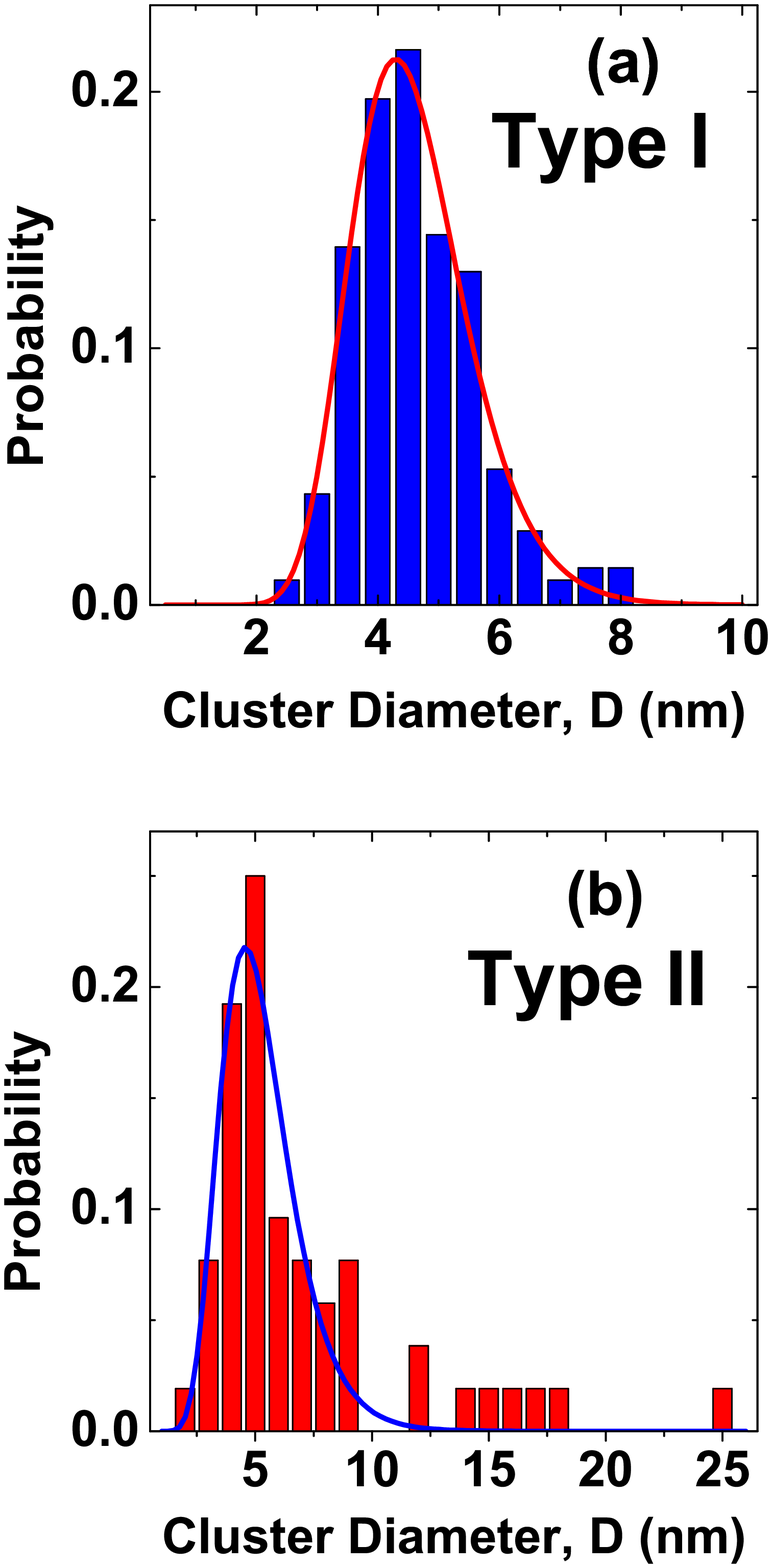}
\caption{Log-normal-fitted cluster size distributions for a Type I and a Type II sample (samples 209B and 203A, respectively).}
\label{fig:SizeDistCompare}
\end{figure}

We now compare our materials to those studied in previous work \cite{Kwiatkowski:2009ys,DiPietro:2010kx} by carrying out a statistical survey of the diameters of these (assumed spherical) nanoclusters in both the Type I material and the Type II material.  Figure \ref{fig:SizeDistCompare} shows the data compiled by measuring the diameters of the clusters in HRXTEM images and plotting the normalized number of clusters versus their diameter $D$.  After this, we applied a log-normal fit to the plot to determine the distribution's width $\sigma$ and the median cluster diameter for the sample $D_0$. In order to perform this fit, we used the relation:\cite{Chen:2010nx}

\begin{equation}
p(D)=\frac{A}{\sigma D\sqrt{2\pi}}\exp \left[-\left(\frac{3 \ln (D/D_0)}{\sigma \sqrt{2}}\right)^2\right]
\label{eqn:logNormal}
\end{equation}
where $\sigma$ is the width or variance of the distribution, $D_0$ is the median diameter of a particle in the material, and $A$ is a scaling factor that normalizes the probability density function $p(D)$.  Applying this fitting function, as shown in Figure \ref{fig:SizeDistCompare}, yielded widths of $0.64\pm 0.02$ and $0.93\pm 0.09$ for the Type I and Type II materials, respectively.  It also yielded median diameters of $4.49\pm 0.03$ nm and $5.03\pm 0.15$ nm for the materials. The values for the median diameters for the Type I and Type II materials are reasonable when compared to the data, indicating a meaningful fit.  This analysis indicates that the distribution of nanoclusters in our samples is of a generically similar nature to those studied earlier, with the notable exception that they exhibit a marked transition from Type I to Type II as a function of Mn content.

\section{Summary}
\label{sec:summary}
 
In summary, we have found that, in a co-doped (Ga,Mn)As:Be system, two distinct materials systems are obtained upon {\it in situ} annealing. For Mn content in the range $5\% - 7.5 \%$, annealing creates a superparamagnetic material with a uniform distribution of small clusters (diameter $\sim 6$ nm) and with a low blocking temperature (\tb $\sim 10$ K). While transmission electron microscopy cannot definitively identify the composition and crystalline phase of the small clusters in this Type I material, our experimental data suggest that they may be comprised of either zinc-blende MnAs or Mn-rich regions of (Ga,Mn)As. At higher Mn content ($\gtrsim 8 \%$), our TEM studies show that annealing results in an inhomogeneous distribution of small clusters and much larger NiAs-phase MnAs clusters (diameter $\sim 25$ nm). Magnetometry reveals that this Type II material also exhibits supermagnetism, albeit with substantially larger magnetic moments, coercive fields and blocking temperatures (well above room temperature). The substantial orientation dependence of the magnetic properties of this material suggests a preferential crystalline orientation to the large MnAs clusters. Finally, we fit the size distribution to a log-normal curve and extracted a median diameter and distribution width for each type of material. 

The synthesis protocols developed in this work provide a promising pathway for designing hybrid ferromagnet/semiconductor nanomaterials for spintronics. In particular, our results indicate that the nanocluster size, distribution and spatial location depend upon several factors including Mn content, Be-doping and vicinal interfaces. Further experiments are currently underway to explore whether useful magnetoresistance or magnetooptical phenomena result from the interplay between the itinerant holes in the p-GaAs host lattice and the MnAs clusters.

\begin{acknowledgments}
The authors acknowledge useful discussions with R. Misra, J. Kulik, A. Balk, and M. Dahlberg about many aspects of this work.  This research is supported by the ONR MURI program under Contract No. N0014-06-1-0428. This work was supported by the Pennsylvania State University Materials Research Institute Nanofabrication Lab and the National Science Foundation Cooperative Agreement No. 0335765, National Nanotechnology Infrastructure Network, with Cornell University.

\end{acknowledgments}


\newpage
\begin{table}[h]
\caption{Table showing the different MBE growth parameters used for nanocluster synthesis.  The 8.1\% column is highlighted with two vertical lines to indicate that these samples are Type II materials, with the rest of the table being Type I materials.  Sample data shown in this work comes from samples 208A and B, 209A and B, and 203A.  Be cell temperatures are given instead of Be concentrations as the Be molecular beam flux is too low for our measurement technique to measure and thus concentrations cannot be calculated.}

\label{table:SampleTable}

  \begin{tabular}{| c || c || c | c | c | c |}
  	\hline
		{\bf Be Cell Temperature ($^ {\circ}$C)} & \multicolumn{5}{|c|}{{\bf Mn Content}} \\
		\hline
		 													 & {\bf 8.1\%} & {\bf 7.5\%} & {\bf 6.8\%} & {\bf 6.2\%} & {\bf 5.5\%} \\ \hline
    \multirow{3}{*}{{\bf 990}} & 214B & 209B & 209A & 208B & 208A \\
    										 			 & 204B & 210A &      &      &      \\
    										 			 & 209C &      &      &      &      \\ \hline
    \multirow{2}{*}{{\bf 970}} & 121B &      &      &      &      \\
    										 			 & 204A &      &      &      &      \\ \hline
    \multirow{2}{*}{{\bf 950}} & 121A &      &      &      &      \\
    										 			 & 203B &      &      &      &      \\ \hline
    								 {\bf 915} & 203A &      &      &      &      \\ \hline
    								 {\bf 890} & 204C &      &      &      &      \\ \hline
  \end{tabular}
\end{table}

\newpage

\newpage

\newpage

\newpage

\newpage

\newpage

\newpage

\newpage

\newpage

\newpage

\newpage

\newpage

\end{document}